\documentclass[pra,aps,eqsecnum,amsmath,amssymb,showkeys]{revtex4}
\newcommand{\am}{\mathsf{A}}
\newcommand{\bn}{\mathsf{B}}
\newcommand{\cn}{\mathsf{C}}
\newcommand{\hh}{\mathcal{H}}
\newcommand{\lnp}{\mathcal{L}}

\newcommand{\ip}{\mathsf{\Pi}}

\newcommand{\bsg}{\boldsymbol{\sigma}}
\newcommand{\bro}{\boldsymbol{\rho}}

\newcommand{\tr}{{\mathrm{tr}}}
\newcommand{\kr}{{\mathrm{ker}}}
\newcommand{\sop}{\mathrm{supp}}
\newcommand{\spc}{\mathrm{spec}}
\newcommand{\ron}{{\mathrm{ran}}}
\newcommand{\rh}{{\mathrm{H}}}
\newcommand{\rd}{{\mathrm{D}}}
\newcommand{\rs}{{\mathrm{S}}}

\newcommand{\mcx}{\Omega}
\newcommand{\llv}{\varLambda}
\newcommand{\rrv}{\varUpsilon}
\newcommand{\ax}{\mathsf{X}}
\newcommand{\ay}{\mathsf{Y}}
\newcommand{\np}{\mathsf{P}}
\newcommand{\nq}{\mathsf{Q}}
\newcommand{\veps}{\varepsilon}
\newcommand{\pen}{\openone}

\newcommand{\lsp}{{\cal{L}}_{+}}
\newcommand{\tom}{\widetilde{\Omega}}
\begin{document}

\preprint{}

\title{{\bf Bounds of the Pinsker and Fannes Types on the Tsallis Relative Entropy}}

\author{Alexey E. Rastegin}
 \affiliation{Department of Theoretical Physics, Irkutsk State University,
Gagarin Bv. 20, Irkutsk 664003, Russia}

\begin{abstract}
Pinsker's and Fannes' type bounds on the Tsallis relative entropy
are derived. The monotonicity property of the quantum
$f$-divergence is used for its estimating from below. For order
$\alpha\in(0,1)$, a family of lower bounds of Pinsker type is
obtained. For $\alpha>1$ and the commutative case, upper
continuity bounds on the relative entropy in terms of the minimal
probability in its second argument are derived. Both the lower and
upper bounds presented are reformulated for the case of
R\'{e}nyi's entropies. The Fano inequality is extended to Tsallis'
entropies for all $\alpha>0$. The deduced bounds on the Tsallis
conditional entropy are used for obtaining inequalities of Fannes' type. 
\end{abstract}


\maketitle

\pagenumbering{arabic}
\setcounter{page}{1}

\section{Introduction}\label{intr}

Physical systems with long-range interactions, long-time memories,
or fractal structures can hardly be treated within the traditional
background of statistical physics. The Tsallis entropies have been
widely adopted in this direction \cite{gmt}. As a rule, stationary
states of such systems are described by one-parametric extensions
of the Zipf-Mandelbrot power-law distribution. Generalized
entropies have also found use as alternate measures of an
informational content. For instance, the entropic uncertainty
principle has been expressed in terms of both the Tsallis
\cite{majer,rast104} and R\'{e}nyi entropies
\cite{rastijtp,zpv08}. Studies of generalized entropies allow to
treat properties of the standard entropy in more general setting.
The connection between strong subadditivity of the von Neumann
entropy and the Wigner-Yanase-Dyson conjecture is a remarkable
example (see \cite{hmp10,JR10} and references therein).

The relative entropy, or Kullback-Leibler divergence \cite{KL51},
is frequently used as a measure of statistical distinguishability.
Csiszar's $f$-divergence \cite{ics67,ics08} and Petz's
quasi-entropies \cite{petz86,petz10} are famous generalizations of
the Kullback-Leibler measure to the classical and quantum cases,
respectively. In both the cases, properties of the relative
entropy are the subject of active research. So, the development of
a standard background to generalized entropies is an important
issue. Lower bounds of the Pinsker type on the classical
$f$-divergences were given by Csisz\'{a}r \cite{ics66}. In this
paper, the function in which the variational distance is
substituted is not explicit. Gilardoni presented explicit lower
bounds on the classical $f$-divergences including the case of
Tsallis' relative entropy \cite{ggil}.

In the present paper, we deduce some lower and upper bounds on the
Tsallis relative entropy. The obtained bounds are expressed in
terms of the trace distance between two probability distributions
or density operators. The paper is organized as follows. In
Section \ref{dfnt}, the main definitions are given. One
consequence of the monotonicity of the quantum $f$-divergence is
considered in Section \ref{acof}. A family of lower bounds on the
Tsallis relative entropy of order $\alpha\in(0,1)$ is derived in
Section \ref{inpt}. In their essence, these inequalities are
one-parametric extensions of the Pinsker inequality. The case of
R\'{e}nyi relative entropy is considered as well. In Section
\ref{upbn}, upper bounds on the Tsallis relative $\alpha$-entropy
of two probability distributions in terms of the minimal
probability in its second argument are obtained. Fano type upper
bounds on the conditional Tsallis entropy are derived for all
$\alpha>0$ in Section \ref{ngfn}. As is shown, these bounds lead
to generalized inequalities of Fannes type.

\section{Definitions and notation}\label{dfnt}

In the classical regime, we will consider probability
distributions over the index set $\mcx$ of finite cardinality $N$.
The trace distance between probability distributions $P=\{p(x)\}$
and $Q=\{q(x)\}$ are then defined as
\begin{equation}
D(P,Q):=\frac{1}{2}{\,}\sum\nolimits_{x\in\mcx}\bigl|p(x)-q(x)\bigr|
\ . \label{pdpq}
\end{equation}
Let $\lnp(\hh)$ be the space of linear operators on
finite-dimensional Hilbert space $\hh$. We also use the notation
$\lsp(\hh)$ to denote the set of positive semidefinite operators.
For any operator $\ax$, we put $|\ax|\in\lsp(\hh)$ as a unique
positive square root of $\ax^{*}\ax\geq{\mathbf{0}}$. The trace
norm $\|\ax\|_1:=\tr|\ax|$ and the trace distance
\begin{equation}
\rd(\ax,\ay):=\frac{1}{2}{\>}\|\ax-\ay\|_{1}\equiv\frac{1}{2}{\>}\tr|\ax-\ay|
\label{rdrs}
\end{equation}
are widely used in both the mathematical physics and quantum
information theory. Using the Ky Fan norms, the partitioned
versions of the above measures can be adopted properly
\cite{rast1023,rast103}. By $\kr(\ax)$ and $\ron(\ax)$ we denote
the kernel and the range of operator $\ax\in\lnp(\hh)$.
Eigenvalues of the operator $\ax$ form the multi-set $\spc(\ax)$.
For $\ax,\ay\in\lnp(\hh)$, we define the Hilbert--Schmidt inner
product by
\begin{equation}
\langle\ax{\,},\ay\rangle_{\rm{hs}}
:={\rm{tr}}(\ax^{*}\ay)
\ . \label{hsdef}
\end{equation}

For positive $\alpha\not=1$, the Tsallis $\alpha$-entropy of
probability distribution $P=\{p(x)\}$ is defined by \cite{tsallis}
\begin{equation}
H_{\alpha}(P):=\frac{1}{1-\alpha}{\,}\left(\sum\nolimits_{x\in\mcx} p(x)^{\alpha}
- 1 \right)\equiv
-\sum\nolimits_{x\in\mcx} p(x)^{\alpha}\ln_{\alpha}p(x)
\ , \label{tsaent}
\end{equation}
where $\ln_{\alpha}z\equiv\bigl(z^{1-\alpha}-1\bigr)/(1-\alpha)$
is the $\alpha$-logarithm. The maximal value $\ln_{\alpha}N$ is
reached by the uniform distribution, when $p(x)=1/N$ for all
$x\in\mcx$. The Shannon entropy $H_{1}(P)=-\sum_{x}p(x)\ln{p}(x)$
is recovered in the limit $\alpha\to1$. By $h_{\alpha}(u)$ we
denote the binary Tsallis $\alpha$-entropy, i.e.
\begin{equation}
h_{\alpha}(u):= H_{\alpha}\bigl(\{u,1-u\}\bigr)=
-u^{\alpha}\ln_{\alpha}u-(1-u)^{\alpha}\ln_{\alpha}(1-u)
\ , \label{hbin}
\end{equation}
where $u\in[0,1]$. This function is clearly concave and obeys
$h_{\alpha}(u)=h_{\alpha}(1-u)$. Other important one-parametric
generalization is the R\'{e}nyi entropy (see \cite{ics08,petz08}):
\begin{equation}
R_{\alpha}(P):=\frac{1}{1-\alpha}{\ }\ln\left(\sum\nolimits_{x\in\mcx} p(x)^{\alpha}\right)
\ . \label{rpdf}
\end{equation}
The obvious formula
$(1-\alpha)R_{\alpha}(P)=\ln\bigl[1+(1-\alpha)H_{\alpha}(P)\bigr]$
relates the entropies (\ref{tsaent}) and (\ref{rpdf}). The quantum
analogs of these entropies are respectively defined as
\begin{align}
{\rm{H}}_{\alpha}(\bro)&:=\frac{1}{1-\alpha}{\>}\Bigl(\tr(\bro^{\alpha})-1\Bigr)
\ , \label{qtsdf}\\
{\rm{R}}_{\alpha}(\bro)&:=\frac{1}{1-\alpha}{\ }\ln\bigl[\tr(\bro^{\alpha})\bigr]
\ . \label{qrndf}
\end{align}
where $\tr(\bro)=1$. Additivity properties of quantum entropies
are an important issue. Subadditivity of the quantum Tsallis
entropy (\ref{qtsdf}) for $\alpha>1$ has been conjectured by
Raggio \cite{raggio} and later proved by Audenaert \cite{auden07}.
This result has been extended to some of so-called unified
entropies \cite{rastjst}. The subadditivity property was believed
to be true for the Wigner-Yanase entropy, until counterexamples
were given \cite{hansen07,seiringer07}. Meantime, it is sufficient
for the subadditivity that the bipartite state is pure. Other
sufficient conditions for subadditivity of the Wigner-Yanase
entropy are obtained in \cite{CH10}.

The standard relative entropy of $P=\{p(x)\}$ to $Q=\{q(x)\}$ is
defined as \cite{ics08}
\begin{equation}
H_{1}(P||Q)=-\sum\nolimits_{x\in\Omega} p(x){\,}\ln\frac{q(x)}{p(x)}
\ . \label{srcd}
\end{equation}
For density operators $\bro$ and $\bsg$, the
quantum relative entropy is expressed as \cite{petz08}
\begin{equation}
\rh_{1}(\bro||\bsg):=
\left\{
\begin{array}{ll}
\tr(\bro\ln\bro-\bro\ln\bsg) \ ,
&\quad {\mathrm{if}}{\ }\sop(\bro)\leq\sop(\bsg) \ , \\
+\infty\ , &\quad {\mathrm{otherwise}} \ .
\end{array}
\right.
\label{relan}
\end{equation}
By $\sop(\am)$, we mean the support projection of
$\am\in\lsp(\hh)$. Instead of the entry
$\sop(\bro)\leq\sop(\bsg)$, the condition
$\kr(\bsg)\subset\kr(\bro)$ can be written as well \cite{ruskai}.
In the classical regime, the Tsallis relative $\alpha$-entropy is
introduced by \cite{borland}
\begin{equation}
H_{\alpha}(P||Q):=-\sum\nolimits_{x\in\mcx}p(x){\,}\ln_{\alpha}\frac{q(x)}{p(x)}=
\frac{1}{1-\alpha}{\,}\left(1-\sum\nolimits_{x\in\mcx}p(x)^{\alpha}q(x)^{1-\alpha} \right)
{\>}. \label{srtdef}
\end{equation}
Basic properties of this measure are discussed in
\cite{borland,fky04}. The R\'{e}nyi relative entropy is defined as
\cite{ics08}
\begin{equation}
R_{\alpha}(P||Q):=\frac{-1}{1-\alpha}{\ }\ln\left(
\sum\nolimits_{x\in\mcx}p(x)^{\alpha}q(x)^{1-\alpha}\right)
{\>}. \label{rrdef}
\end{equation}
Gilardoni derived Pinsker's type inequalities for both the Tsallis
and R\'{e}nyi relative entropies \cite{ggil}. Note that the
Tsallis relative entropy is written in \cite{ggil} with the
denominator $\alpha(1-\alpha)$ instead of $(1-\alpha)$.

Let us to extend the definition (\ref{srtdef}) to any
positive-valued functions $A$ and $B$ on the finite set $\mcx$.
For given set $A=\{a(x)\}$, we put the index subset
$\mcx_A=\{x:{\,}a(x)\neq0\}$ and its complement $\tom_A$. For
$\alpha>1$, the ''Tsallis relative $\alpha$-entropy'' of
$A=\{a(x)\}$ to $B=\{b(x)\}$ is defined as
\begin{equation}
H_{\alpha}(A||B):=
\left\{
\begin{array}{ll}
\frac{1}{\alpha-1}\Bigl(\sum\nolimits_{x\in\mcx_A}
a(x)^{\alpha}b(x)^{1-\alpha}-\sum\nolimits_{x\in\mcx_A} a(x)\Bigr)\>,
& {\ }{\ } \Omega_{A}\subset\Omega_{B} \ , \\
+\infty\ , & {\ }{\ } {\rm{otherwise}} \ .
\end{array}
\right.
\label{geendf}
\end{equation}
Omitting the second entry, we obtain the definition for
$0<\alpha<1$. For any positive scalar $\lambda$, we have
\begin{equation}
H_{\alpha}(\lambda{A}||\lambda{B})=\lambda{\,}H_{\alpha}(A||B)
\ , \label{homg}
\end{equation}
i.e. it is a homogeneous function of degree one. For
$\alpha\in(0,1)$ and density operators $\bro$ and $\bsg$, we
define the Tsallis relative entropy as
\begin{equation}
\rh_{\alpha}(\bro||\bsg):=\frac{1}{1-\alpha}
{\ }\left(1-\tr\bigl(\bro^{\alpha}\bsg^{1-\alpha}\bigl)\right)
\ . \label{ahadef}
\end{equation}
For $\alpha>1$, the right-hand side of (\ref{ahadef}) is
well-defined whenever $\sop(\bro)\leq\sop(\bsg)$. In the singular
case, when $\sop(\bro)\nleq\sop(\bsg)$, the right-hand side of
(\ref{ahadef}) is dealt similar to the standard relative entropy
(\ref{relan}). Namely, relative entropies are defined to be
$+\infty$. Extending (\ref{rrdef}) to the quantum case, we define
\begin{equation}
{\rm{R}}_{\alpha}(\bro||\bsg):=\frac{-1}{1-\alpha}{\ }\ln\bigl[\tr(\bro^{\alpha}\bsg^{1-\alpha})\bigr]
\ . \label{radef}
\end{equation}
For these entropies, we have the formula
\begin{equation}
(\alpha-1){\rm{R}}_{\alpha}(\bro||\bsg)=\ln\bigl[1+(\alpha-1)\rh_{\alpha}(\bro||\bsg)\bigr]
\ , \label{alrha}
\end{equation}
and its classical variety. For $\alpha>1$ and
$\am,\bn\in{\mathcal{L}}_{+}(\hh)$, we also introduce
\begin{equation}
\rh_{\alpha}(\am||\bn):=
\left\{
\begin{array}{ll}
\frac{1}{\alpha-1}{\,}\Bigl(\tr(\am^{\alpha}\bn^{1-\alpha})-\tr(\am)\Bigr)\ ,
& {\ }{\ } \ron(\am)\subset\ron(\bn) {\ ,} \\
+\infty\ , & {\ }{\ } {\rm{otherwise}} \ .
\end{array}
\right.
\label{qendf}
\end{equation}

\section{A consequence of monotonicity of the $f$-divergence}\label{acof}

In this section, we discuss one result which will be used to
obtain quantum bounds of Pinsker's type. The Tsallis relative
entropy (\ref{srtdef}) is closely related to the Csisz\'{a}r
$f$-divergence \cite{ics67}. Let $z\mapsto{f}(z)$ be a convex
function on $z\in[0,+\infty)$ with $f(1)=0$. The Csisz\'{a}r
$f$-divergence of $P=\{p(x)\}$ from $Q=\{q(x)\}$ is defined as
\cite{ics67,ics08}
\begin{equation}
S_{f}(P||Q):=\sum_{x\in\mcx} q(x){\>}f{\!}\left(\frac{p(x)}{q(x)}\right)
\ . \label{cfdf}
\end{equation}
Taking $f_{\alpha}(z)=\bigl(z^{\alpha}-z\bigr)/(\alpha-1)$ with
positive $\alpha\neq1$, the formula (\ref{cfdf}) leads to
(\ref{srtdef}). The standard case is recovered with
$f(z)=z\ln{z}$. The definition (\ref{cfdf}) can generally be used
without the normalization condition.

In the following, we use the convention that powers of a positive
semidefinite operator are taken only on its support. So, by
$\am^{-1}$ and $\am^{0}$ we respectively mean the generalized
inverse and the support projection of $\am$. A quantum counterpart
of Csisz\'{a}r's $f$-divergence is introduced as follows
\cite{hmp10}. For an operator $\am\in\mathcal{L}_{+}(\hh)$, let
$\llv_{\am}$ and $\rrv_{\am}$ denote the left and the right
multiplications by $\am$, respectively, defined as
\begin{equation}
\llv_{\am}:{\>} \ax\mapsto\am\ax\ , \qquad
\rrv_{\am}:{\>} \ax\mapsto\ax\am \ , \qquad \ax\in\mathcal{L}(\hh)
\ . \label{lrmults}
\end{equation}
Left and right multiplications commute with each other, namely
$\llv_{\am}\rrv_{\bn}=\rrv_{\bn}\llv_{\am}$ for
$\am,\bn\in\mathcal{L}_{+}(\hh)$. Let $z\mapsto{f}(z)$ be a
continuous function on $z\in[0,+\infty)$. Taking the set
$\bigl\{ab^{-1}:{\>}a\in\spc(\am),{\>}b\in\spc(\bn)\bigr\}$, we
write \cite{hmp10}
\begin{equation}
f(\llv_{\am}\rrv_{\bn^{-1}}):=\sum_{a\in\spc(\am)}{\,}\sum_{b\in\spc(\bn)}
f(ab^{-1}){\,}\llv_{\np_{a}}{\,}\rrv_{\nq_{b}}
\ , \label{fabpq}
\end{equation}
where the formulas $\am=\sum_{a}a{\,}\np_{a}$ and
$\bn=\sum_{b}b{\,}\nq_{b}$ respectively express the spectral
decompositions of $\am$ and $\bn$. If $\ron(\am)\subset\ron(\bn)$,
then the $f$-divergence of $\am$ with respect to $\bn$ is defined
as \cite{hmp10}
\begin{equation}
\rs_{f}(\am||\bn):=\bigl\langle\bn^{1/2},f(\llv_{\am}\rrv_{\bn^{-1}}){\,}\bn^{1/2}\bigr\rangle_{\rm{hs}}
\ . \label{qsfdef}
\end{equation}
Let $\pen$ be the identity operator. In general case, the quantum
$f$-divergence is defined by the formula
\begin{equation}
\rs_{f}(\am||\bn):=\underset{\veps\searrow{0}}{\lim}{\>}\rs_{f}(\am||\bn+\veps\pen)
\ . \label{qsfdep}
\end{equation}
Basic properties of the quantity (\ref{qsfdef}) are discussed in
the paper \cite{hmp10}. Using the function
$f_{\alpha}(z)=\bigl(z^{\alpha}-z\bigr)/(\alpha-1)$, we actually
obtain the quantity (\ref{qendf}). One of the most important
properties of relative entropies is their monotonicity under the
action of trace-preserving completely positive (TPCP) maps
\cite{uhlmann77}. For general discussion of a role of stochastic
maps in quantum theory, see the paper \cite{busch99}. Many
fundamental results of quantum information theory are closely
related to the monotonicity of the standard relative entropy
\cite{JR10,nielsen,vedral02}. General conditions for the
monotonicity of the quantum $f$-divergence are obtained in
\cite{hmp10}. If the map $\varPhi$ is TPCP-map and the function
$f$ is operator convex on $[0,+\infty)$ then
\begin{equation}
\rs_{f}\bigl(\varPhi(\am)\big|\big|{\,}\varPhi(\bn)\bigr)\leq\rs_{f}(\am||\bn)
\ . \label{fdmon}
\end{equation}
Note that the inequality (\ref{fdmon}) has generally been
established in \cite{hmp10} under weaker conditions on the maps.
From the monotonicity (\ref{fdmon}) we can derive simple upper
bounds on the quantum $f$-divergence in terms of classical one.
Let $\ip\in\lsp(\hh)$ be a projection. It is known that the linear
map
\begin{equation}
\ax\mapsto\bigl\{\tr(\ip\ax),\tr[(\pen-\ip)\ax]\bigr\}
\label{ipmp}
\end{equation}
is both trace-preserving and completely positive. Combining this
fact with the monotonicity (\ref{fdmon}), we have arrived at a
conclusion.

\newtheorem{lem1}{Lemma}
\begin{lem1}\label{conmon1}
Suppose $\am,\bn\in\lsp(\hh)$. Let $\ip_{\pm}\in\lsp(\hh)$ obey
$\ip_{+}+\ip_{-}={\textup{\pen}}$ and be projectors onto the
eigenspaces corresponding to positive and negative eigenvalues of
$(\am-\bn)$, respectively. If the function $f$ is operator convex
then
\begin{equation}
\rs_{f}(\am||\bn)\geq{S}_{f}\bigl(\{u'_{\pm}\}\big|\big|\{v'_{\pm}\}\bigr)
\ , \label{sfuf}
\end{equation}
where $u'_{\pm}=\tr(\ip_{\pm}\am)$ and $v'_{\pm}=\tr(\ip_{\pm}\bn)$.
\end{lem1}

The function $z\mapsto{z}^{\alpha}$ is operator concave on
$\lsp(\hh)$ for $0\leq\alpha\leq1$ and operator convex on
$\lsp(\hh)$ for $1\leq\alpha\leq2$ (see, respectively, theorems
4.2.3 and 1.5.8 in \cite{bhatia07}). So the function
$f_{\alpha}(z)=\bigl(z^{\alpha}-z\bigr)/(\alpha-1)$ is operator
convex for $\alpha\in[0,2]$ and $\alpha\neq1$. Combining this with
the inequality (\ref{sfuf}) gives
\begin{equation}
\rh_{\alpha}(\am||\bn)\geq{H}_{\alpha}\bigl(\{u'_{\pm}\}\big|\big|\{v'_{\pm}\}\bigr)
\ . \label{pinab}
\end{equation}
Up to a notation, the result (\ref{pinab}) with density operators
was presented in \cite{rast10fn} (see theorem IV.1 therein). In
the next section, we will use the relations (\ref{pinab}) and
\begin{equation}
\|\am-\bn\|_{1}=|u'_{+}-v'_{+}|+|u'_{-}-v'_{-}|
\label{abtrn}
\end{equation}
for estimating the right-hand of (\ref{sfuf}) from below in terms
of the distance $\|\am-\bn\|_{1}$.

\section{Pinsker type inequalities for $\alpha\in(0,1)$}\label{inpt}

Studies of distinguishability measures and relations between them
is an actual issue of quantum information theory. The Pinsker
inequality \cite{ics67} and its quantum analog expressed as
\cite{hots81}
\begin{equation}
\rh_1(\bro||\bsg)\geq2{\,}\rd(\bro,\bsg)^2
\ , \label{qpins}
\end{equation}
are well-known results of such a kind. Lower and upper bounds on
the relative entropy (\ref{relan}) were given in \cite{AE05}.
Upper bounds of the papers \cite{AE05,AE11} are similar to Fannes'
inequality \cite{fannes}. The authors of \cite{still90} proved
that
\begin{equation}
\rh_{\alpha}(\bro||\bsg)\leq\rh_1(\bro||\bsg)\leq\rh_{\beta}(\bro||\bsg)
\ , \label{p1qin}
\end{equation}
where $0\leq\alpha<1$ and $1<\beta\leq2$. So, for $1<\beta\leq2$
the relative entropy $\rh_{\beta}(\bro||\bsg)$ is bounded from
below by the right-hand side of Eq. (\ref{qpins}). More detailed
lower bounds on the relative entropy (\ref{relan}) are presented
in \cite{AE05}. By (\ref{p1qin}), these lower bounds are all
valid for $\rh_{\beta}(\bro||\bsg)$ with $1<\beta\leq2$.

Let $\ip_{+}$ be a projector on the eigenspace corresponding to
positive eigenvalues of the difference $(\bro-\bsg)$. For
normalized density operators, the inequality (\ref{pinab})
together with the definitions (\ref{srtdef}) and (\ref{ahadef})
leads to the bound
\begin{equation}
\rh_{\alpha}(\bro||\bsg)\geq{H}_{\alpha}\bigl(\{u,1-u\}\big|\big|\{v,1-v\}\bigr)
\ , \label{piniq}
\end{equation}
where we write $u=\tr(\ip_{+}\bro)$ and $v=\tr(\ip_{+}\bsg)$ for
brevity. Denoting $t=|u-v|$, we also have $\|\bro-\bsg\|_{1}=2t$
and $\rd(\bro,\bsg)=t$. In the paper \cite{rast10fn}, for
$u,v\in[0,1]$ we have proved the inequality (see lemma IV.2
therein)
\begin{equation}
\sqrt{uv}+\sqrt{(1-u)(1-v)}\leq\sqrt{1-t^2}
\ , \label{tih12}
\end{equation}
whence
${H}_{1/2}\bigl(\{u,1-u\}\big|\big|\{v,1-v\}\bigr)\geq2\bigl(1-\sqrt{1-t^2}\bigr)$
and $\rh_{1/2}(\bro||\bsg)\geq\rd(\bro,\bsg)^2$. Note that the
result (\ref{tih12}) is actually a special case of the inequality
\begin{equation}
\rd(\bro,\bsg)\leq\sqrt{1-F(\bro,\bsg)^{2}}
\ , \label{fvdg}
\end{equation}
where $F(\bro,\bsg)=\tr\bigl|\sqrt{\bro}\sqrt{\bsg}\bigr|$ is the
fidelity of $\bro$ and $\bsg$. The formula (\ref{fvdg}) was proved
by Fuchs and van der Graaf \cite{graaf}. Using (\ref{tih12}), we
obtain the following statement.

\newtheorem{lem2}[lem1]{Lemma}
\begin{lem2}\label{conal}
Let $u,v\in[0,1]$ and $g(t)=1-\sqrt{1-t^2}$. For
$\alpha\in[0,1/2]$ and $t=|u-v|$, there holds
\begin{equation}
u^{\alpha}v^{1-\alpha}+(1-u)^{\alpha}(1-v)^{1-\alpha}\leq1-2\alpha{g}(t)
\ . \label{alcon}
\end{equation}
\end{lem2}

{\bf Proof.} For fixed $u$ and $v$, we define the function
\begin{equation}
\Phi_{uv}(\alpha)=u^{\alpha}v^{1-\alpha}+(1-u)^{\alpha}(1-v)^{1-\alpha}+2\alpha{g}(t)-1
\ . \label{alcon1}
\end{equation}
The claim (\ref{alcon}) is equivalent to the inequality
$\Phi_{uv}(\alpha)\leq0$ for all $\alpha\in[0,1/2]$. First, we
have $\Phi_{uv}(0)\leq0$ obviously; second, $\Phi_{uv}(1/2)\leq0$
because of (\ref{tih12}). Third, $\Phi_{uv}(\alpha)$ is a convex
function of the parameter $\alpha$. Indeed, for $u,v\neq0,1$ we
write down
\begin{equation}
\frac{\partial^2\Phi_{uv}}{\partial\alpha^2}=u^{\alpha}v^{1-\alpha}\left(\ln\frac{u}{v}\right)^2+
(1-u)^{\alpha}(1-v)^{1-\alpha}\left(\ln\frac{1-u}{1-v}\right)^2\geq0
\ . \label{parph}
\end{equation}
If a convex function is negative at the end points of some
interval, it is negative in this interval everywhere.
$\blacksquare$

Combining the statement of Lemma \ref{conal} with (\ref{pinab})
gives a lower bound of the Pinsker type on the Tsallis relative
$\alpha$-entropy for $\alpha\in(0,1)$. We formulate it for two
positive operators with equal traces.

\newtheorem{tem1}{Theorem}
\begin{tem1}\label{bbpn}
Let $\am,\bn\in\lsp(\hh)$, $\tr(\am)=\tr(\bn)=\theta$,
$\rd(\am,\bn)=\tau$ and $g(t)=1-\sqrt{1-t^2}$. For all
$\alpha\in(0,1)$, there holds
\begin{equation}
\rh_{\alpha}(\am||\bn)\geq\varkappa_{\alpha}{\,}\theta{\,}{g}\bigl(\tau/\theta\bigr)
\ , \label{rhavf}
\end{equation}
where the factor $\varkappa_{\alpha}=2\alpha(1-\alpha)^{-1}$ for
$0<\alpha\leq1/2$ and $\varkappa_{\alpha}=2$ for
$1/2\leq\alpha<1$.
\end{tem1}

{\bf Proof.} Using the notation of Lemma \ref{conmon1}, we
have $u'_{+}+u'_{-}=v'_{+}+v'_{-}=\theta$ and
$\|\am-\bn\|_{1}=2|u'_{+}-v'_{+}|$, whence $\tau=|u'_{+}-v'_{+}|$.
It follows from (\ref{pinab}) and (\ref{geendf}) that
\begin{align}
(1-\alpha){\,}\rh_{\alpha}(\am||\bn)
&\geq(1-\alpha){\,}H_{\alpha}\bigl(\{u'_{\pm}\}\big|\big|\{v'_{\pm}\}\bigr)
\nonumber\\
&=\theta\left[1-u^{\alpha}v^{1-\alpha}-(1-u)^{\alpha}(1-v)^{1-\alpha}\right]
\ , \label{al12}
\end{align}
where $u=u'_{+}/\theta$, $v=v'_{+}/\theta$, and
$|u-v|=\tau/\theta$. Due to (\ref{alcon}), for $0<\alpha\leq1/2$
the right-hand side of (\ref{al12}) is not less than
$2\alpha{\,}\theta{\,}g(\tau/\theta)$. Hence the claim
(\ref{rhavf}) with $\varkappa_{\alpha}=2\alpha(1-\alpha)^{-1}$
follows. For $1/2\leq\alpha<1$, we put $\beta=1-\alpha$ and
further write
\begin{equation}
\beta{\,}\rh_{\alpha}(\bro||\bsg)\geq
\theta\left[1-u^{1-\beta}v^{\beta}-(1-u)^{1-\beta}(1-v)^{\beta}\right]\geq2\beta{\,}\theta{\,}g(\tau/\theta)
\ . \label{al21}
\end{equation}
Hence the claim (\ref{rhavf}) with $\varkappa_{\alpha}=2$ follows.
$\blacksquare$

For probability distributions, the lower bound (\ref{rhavf}) is
rewritten with the classical trace distance $\tau=D(P,Q)$.
Expanding the function $g(\tau/\theta)$ into power series, we
obtain a family of lower bounds of the Pinsker type. Namely, we
have the bound
\begin{equation}
\rh_{\alpha}(\am||\bn)\geq\varkappa_{\alpha}\sum_{n=1}^{\infty}
\binom{1/2}{n}(-1)^{n+1}\frac{\tau^{2n}}{\theta^{2n-1}}
\ , \label{exbn}
\end{equation}
including
$\rh_{\alpha}(\am||\bn)\geq(2\theta)^{-1}\varkappa_{\alpha}\tau^{2}$.
The coefficient $\binom{1/2}{n}(-1)^{n+1}$ is positive for all
$n$. So, any partial sum of the series (\ref{exbn}) provides a
lower bound. In general, this series does not provide an expansion
with the best constants at powers of the trace distance. For the
standard relative entropy $H_{1}(P||Q)$, such constants have been
the subject of long-time research (see \cite{FHT03} and references
therein). Using
$f(z)=\bigl(z^{\alpha}-z\bigr)\big/\bigl(\alpha(\alpha-1)\bigr)$,
Gilardoni \cite{ggil} obtained Pinsker's bound on the
$f$-divergence (\ref{cfdf}) for $\alpha\in[-1,2]$ and
$\alpha\neq0,1$. When $\alpha\in[0,2]$, the results of the paper
\cite{ggil} can directly be combined with (\ref{sfuf}), since the
quantum relative $\alpha$-entropy is monotone here. In our
notation for the Tsallis case, the Pinsker type bound of
\cite{ggil} reads
\begin{equation}
\rh_{\alpha}(\bro||\bsg)\geq2\alpha\tau^{2}+\frac{2}{9}{\,}\alpha(\alpha+1)(2-\alpha)\tau^{4}
\ , \label{sobp}
\end{equation}
where $\tau=\rd(\bro,\bsg)$ is the trace distance between
normalized density matrices. It must be stressed
that the inequality (\ref{rhavf}) is better in some joint region
of the trace distance and $\alpha$ close to 1/2. Say, for $\alpha=1/2$ the
bound (\ref{rhavf}) is stronger for all $\tau\neq0$. In this case,
we obtain from (\ref{exbn}) that
\begin{equation}
\rh_{1/2}(\bro||\bsg)\geq\tau^{2}+\frac{1}{4}{\>}\tau^{4}+
\sum_{n=3}^{\infty}
\binom{1/2}{n}(-1)^{n+1}\tau^{2n}
\ , \label{exbn1}
\end{equation}
whereas the bound (\ref{sobp}) involves only the first two terms
of the right-hand side of (\ref{exbn1}).

For $\alpha\in(0,1)$, we can also combine (\ref{alrha}) with
(\ref{rhavf}) and hence obtain
\begin{equation}
{\rm{R}}_{\alpha}(\bro||\bsg)\geq\frac{1}{\alpha-1}{\ }
\ln\bigl[1-(1-\alpha)\varkappa_{\alpha}g(\tau)\bigr]\geq\varkappa_{\alpha}g(\tau)
\ , \label{hravf}
\end{equation}
where $\tau=\rd(\bro,\bsg)$. Indeed, the function
$x\mapsto(\alpha-1)^{-1}\ln\bigl[1-(1-\alpha){\,}x\bigr]$
increases with $x\in\bigl[0,\frac{1}{1-\alpha}\bigr)$ and
$-\ln(1-\xi)\geq\xi$ for $\xi\in[0,1)$. The inequality
(\ref{hravf}) can be regarded as the bound of Pinsker's type on
the R\'{e}nyi relative entropy for $\alpha\in(0,1)$. Thus, we have
obtained a family of lower bounds in terms of the trace
distance on both the relative entropies (\ref{ahadef}) and
(\ref{radef}).

\section{Upper continuity bounds for $\alpha>1$}\label{upbn}

One of basic features of the standard relative entropy is its
unboundedness. The relative $\alpha$-entropy enjoys the same for
$\alpha>1$. So we may ask a behaviour of $H_{\alpha}(P||Q)$ as the
minimal probability in $Q$ goes to zero. Of course, in the quantum
case this question is more difficult due to the non-commutativity.
For the standard relative entropy, such an upper bound was
obtained in \cite{BR2}, and more bounds were given in
\cite{AE05,AE11}. For the quantum relative $\alpha$-entropy of
order $\alpha>1$, upper bounds in terms of the minimal eigenvalue
of its second entry were obtained in \cite{rast1009}. It turns out
that in the commutative case these bounds can be sharpened
significantly. Our derivation will mainly based on the joint
convexity. Namely, for each positive $\alpha\neq1$ the quantity
(\ref{geendf}) satisfies
\begin{equation}
H_{\alpha}\left(\theta{A}^{\prime}+(1-\theta)A^{\prime\prime}\bigm|\bigm|\theta{B}^{\prime}+(1-\theta)B^{\prime\prime}\right)
\leq\theta
{H}_{\alpha}(A^{\prime}||B^{\prime})+(1-\theta)H_{\alpha}(A^{\prime\prime}||B^{\prime\prime})
{\>}, \label{tab12}
\end{equation}
for all $0\leq\theta\leq1$. This relation follows from the
so-called ''generalized log-sum inequality'' (see (16) in
\cite{borland}). The properties (\ref{homg}) and (\ref{tab12})
lead to the following statement.

\newtheorem{lem3}[lem1]{Lemma}
\begin{lem3}\label{baah}
Let $A$, $B$, $C$ be three sets of positive numbers, and let $\am$,
$\bn$, and $\cn$ be three positive operators. There holds
\begin{align}
& H_{\alpha}(A+C||B+C)\leq{H}_{\alpha}(A||B) & (0\leq\alpha<\infty)
\ , \label{haabc}\\
& \rh_{\alpha}(\am+\cn||\bn+\cn)\leq\rh_{\alpha}(\am||\bn) & (0\leq\alpha\leq2)
\ . \label{haab2}
\end{align}
\end{lem3}

{\bf Proof.} Using (\ref{homg}) and (\ref{tab12}), we merely write
\begin{align}
H_{\alpha}(A+C||B+C)&=2{\,}H_{\alpha}\bigl((A+C)/2\bigm|\bigm|(B+C)/2\bigr)
\nonumber\\
&\leq{H}_{\alpha}(A||B)+H_{\alpha}(C||C)=H_{\alpha}(A||B)
\ , \label{laah}
\end{align}
since $H_{\alpha}(C||C)=0$. The quantum relative
entropy (\ref{qendf}) also enjoys both the homogeneity of degree
one and the joint convexity, but the latter only for
$0\leq\alpha\leq2$ (see, e.g., the review \cite{JR10}). Rewriting
the above arguments with the quantum relative $\alpha$-entropy
instead of the classical one, we have arrived at the claim
(\ref{haab2}). $\blacksquare$

For the standard relative entropy (\ref{relan}), the relation
(\ref{haab2}) was proved in \cite{AE05}. The inequality
(\ref{haabc}) can be utilized to obtain an upper bound on
$H_{\alpha}(P||Q)$ in terms of the trace distance $D(P,Q)$ and the
minimal probability
\begin{equation}
q_{0}:=\min\{q_{j}:{\>}j\in\Omega_P\}
\ . \label{qmdf}
\end{equation}
Here we apply that any sum in $H_{\alpha}(P||Q)$ is effectively
restricted to the index subset $\Omega_P$. Defining the set
$\Delta=P-Q$ with elements $\delta_{j}=p_{j}-q_{j}$, we put
another set $\Bar{Q}$ with positive elements
\begin{equation}
\Bar{q}_{j}:=\max\{q_{0},-\delta_{j}\}
\ . \label{bqdf}
\end{equation}
Writing $Q=\Bar{Q}+(Q-\Bar{Q})$ and using the property
(\ref{haabc}) with $C=Q-\Bar{Q}$, we obtain
\begin{equation}
H_{\alpha}(P||Q)=H_{\alpha}\left(\Delta+\Bar{Q}+(Q-\Bar{Q})\bigm|\bigm|\Bar{Q}+(Q-\Bar{Q})\right)\leq
H_{\alpha}(\Delta+\Bar{Q}{\,}||\Bar{Q})
\ . \label{baq}
\end{equation}
Use of (\ref{haabc}) is correct here by positivity of both
$\Delta+\Bar{Q}$ and $Q-\Bar{Q}$. Indeed, we have
$\delta_{j}+\max\{q_{0},-\delta_{j}\}\geq0$ and
$q_{j}-\max\{q_{0},q_{j}-p_{j}\}\geq0$ due to (\ref{qmdf}). The
maximization of the right-hand side of (\ref{baq}) is under the
conditions $\sum_{j}\delta_{j}=0$ and
$\sum_{j}|\delta_{j}|=2{\,}D(P,Q)$. We separately consider the two
cases, $D(P,Q)\leq{q}_{0}$ and $q_{0}<D(P,Q)\leq1-q_{0}$.

\newtheorem{tem2}[tem1]{Theorem}
\begin{tem2}\label{thal1}
Let $q_{0}$ be defined by (\ref{qmdf}),
$\Omega_{P}\subset\Omega_{Q}$ and $\tau=D(P,Q)$.
For $\alpha>1$, the Tsallis relative $\alpha$-entropy is bounded
from above as
\begin{align}
& H_{\alpha}(P||Q)\leq\frac{1}{\alpha-1}
{\,}\Bigl(
(q_{0}+\tau)^{\alpha}q_{0}^{1-\alpha}+(q_{0}-\tau)^{\alpha}q_{0}^{1-\alpha}-2q_{0}
\Bigr)
& (\tau\leq{q}_{0})
\ , \label{obon1}\\
& H_{\alpha}(P||Q)\leq\frac{1}{\alpha-1}
{\,}\Bigl(
(q_{0}+\tau)^{\alpha}q_{0}^{1-\alpha}-(q_{0}+\tau)
\Bigr)
 & (q_{0}<\tau\leq1-q_{0})
\ . \label{obon2}
\end{align}
\end{tem2}

{\bf Proof.} It is convenient to define three subsets of the set
$\Omega_{P}$ of cardinality $n$:
\begin{align}
\omega_{x}&:=\left\{j:{\>}j\in\Omega_{P},{\>}0<\delta_{j}\right\}
{\>}, \label{omx} \\
\omega_{y}&:=\left\{j:{\>}j\in\Omega_{P},{\>}-q_{0}\leq\delta_{j}<0\right\}
{\>}, \label{omy} \\
\omega_{z}&:=\left\{j:{\>}j\in\Omega_{P},{\>}\delta_{j}<-q_{0}\right\}
{\>}. \label{omz}
\end{align}
We also introduce corresponding $n$-dimensional positive vectors
$X$, $Y$, and $Z$. These vectors respectively have elements
defined as
\begin{equation}
x_{i}:=
\left\{
\begin{array}{cc}
\delta_{i}{\,}, & i\in\omega_{x} \\
0{\,}, & i\notin\omega_{x}
\end{array}
\right.
{\>}, \qquad
y_{j}:=
\left\{
\begin{array}{cc}
-\delta_{j}{\,}, & j\in\omega_{y} \\
0{\,}, & j\notin\omega_{y}
\end{array}
\right.
{\>}, \qquad
z_{k}:=
\left\{
\begin{array}{cc}
-\delta_{k}{\,}, & k\in\omega_{z} \\
0{\,}, & k\notin\omega_{z}
\end{array}
\right.
{\>}. \label{vexyz}
\end{equation}
Hence we write $\Delta=X-Y-Z$. By (\ref{bqdf}), there holds
\begin{equation}
\Bar{q}_{j}=
\left\{
\begin{array}{cc}
q_{0}{\,}, & j\in\omega_{x}\cup\omega_{y} \\
z_{j}{\,}, & j\in\omega_{z}
\end{array}
\right.
{\>}. \label{baqr}
\end{equation}

Let us begin with the case $\tau\leq q_{0}$. Because of
$|\delta_{j}|\leq\tau$, the set $\omega_{z}$ is empty here. We
first assume that the numbers $\delta_{j}$ are all non-zero,
whence $\Omega_{P}=\omega_{x}\cup\omega_{y}$. The conditions
$\sum_{j}\delta_{j}=0$ and $\sum_{j}|\delta_{j}|=2\tau$ are
rewritten as
\begin{equation}
\sum\nolimits_{i\in\omega_{x}}x_{i}=\sum\nolimits_{j\in\omega_{y}}y_{j}=\tau
\ . \label{xycond}
\end{equation}
In terms of $x_{i}$ and $y_{j}$, the right-hand side of
(\ref{baq}) is represented as the function
\begin{equation}
F(x_{i},y_{j})=\frac{1}{\alpha-1}{\,}\left(
\sum\nolimits_{i\in\omega_{x}}(q_{0}+x_{i})^{\alpha}q_{0}^{1-\alpha}+
\sum\nolimits_{j\in\omega_{y}}(q_{0}-y_{j})^{\alpha}q_{0}^{1-\alpha}
-n{q}_{0}
\right)
{\,}. \label{ffun}
\end{equation}
Possible values of the variables $x_{i}$ and $y_{j}$ correspond to
interior points of the simplex defined by the conditions
$0\leq{x}_{i}$, $0\leq{y}_{j}$ and (\ref{xycond}). Recall that the
global maximum of a convex function relative to a convex set is
reached at one of the extreme points of that set \cite{rockaf}.
Hence the maximal value of $F(x_{i},y_{j})$ on the simplex is
equal to the right-hand side of (\ref{obon1}). It is reached when
one of the $x_{i}$'s and one of the $y_{j}$'s are equal to $\tau$
and other are all zero. Of course, values of $F(x_{i},y_{j})$ in
the interior points of the simplex do not exceed this maximum. If
some of the $\delta_{j}$'s are zero then the question is actually
reduced to the above reasons, but with diminished $n$.

In the case $q_{0}<\tau$, we suppose again that the numbers
$\delta_{j}$ are all non-zero. Instead of (\ref{xycond}), we have
\begin{equation}
\sum\nolimits_{i\in\omega_{x}}x_{i}=\sum\nolimits_{j\in\omega_{y}}y_{j}+\sum\nolimits_{k\in\omega_{z}}z_{k}=\tau
\ . \label{yzcond}
\end{equation}
We first modify the right-hand side of (\ref{baq}). From
(\ref{baqr}), we have $\Bar{Q}=q_{0}(I_{x}+I_{y})+Z$ and
\begin{equation}
\Delta+\Bar{Q}=q_{0}(I_{x}+I_{y})+X-Y
\ . \label{dbdb}
\end{equation}
Here $I_{x}$ denote the indicator of the set $\omega_{x}$ taking
the value 1 for $j\in\omega_{x}$ and 0 for $j\notin\omega_{x}$.
Note that if the set $\Omega_Z$ does not intersect with both the
$\Omega_A$ and $\Omega_B$ then
$H_{\alpha}(A||B+Z)=H_{\alpha}(A||B)$. Using this fact twice and
the inequality (\ref{haabc}) again with positive $C=q_{0}I_{y}-Y$,
we rewrite the right-hand side of (\ref{baq}) as
\begin{align}
& H_{\alpha}\Bigl(q_0(I_{x}+I_{y})+X-Y\bigm|\bigm| q_0(I_x+I_y)+Z\Bigr)
=H_{\alpha}\Bigl(q_{0}I_{x}+X+C\bigm|\bigm|{q}_{0}I_{x}+Y+C\Bigr)
\nonumber\\
&\leq{H}_{\alpha}\left(q_{0}I_{x}+X\bigm|\bigm|{q}_{0}I_{x}+Y\right)
={H}_{\alpha}\left(q_{0}I_{x}+X\bigm|\bigm|{q}_{0}I_{x}\right)
{\>}. \label{zcag}
\end{align}
The latter can be rewritten as the function
\begin{equation}
G(x_{i})=\frac{1}{\alpha-1}\left(\sum\nolimits_{i\in\omega_{x}}(q_{0}+x_{i})^{\alpha}q_{0}^{1-\alpha}-(n_{x}q_{0}+\tau)\right)
{\>}, \label{gfun}
\end{equation}
where $n_{x}$ is cardinality of the $\omega_{x}$. Possible values
of the variables $x_{i}$ relate to interior points of the simplex
defined by $0\leq{x}_{i}$ and $\sum_{i}x_{i}=\tau$. So the maximum
of $G(x_{i})$ is equal to the right-hand side of (\ref{obon2}) and
reached, when one of the $x_{i}$'s is $\tau$ and other are all
zero. As above, we reduce the case, in which some of the
$\delta_{j}$'s are zero. $\blacksquare$

The upper bounds (\ref{obon1}) and (\ref{obon2}) have a behavior
$q_{0}^{1-\alpha}$ with respect to the minimal probability
$q_{0}$. For the quantum relative entropy
$\rh_{\alpha}(\bro||\bsg)$, upper bounds with a similar dependence
on the minimal eigenvalue of $\bsg$ were obtained in
\cite{rast1009}. The bounds (\ref{obon1}) and (\ref{obon2}) are
stronger, but their proof is quite restricted to the commutative
case. The principal point is that positivity of diagonal elements
of a matrix do not imply positivity of matrix itself (except for
the case of diagonal matrices). Note that the inequalities
(\ref{obon1}) and (\ref{obon2}) can be rewritten in terms of
$\alpha$-logarithm as
\begin{align}
H_{\alpha}(P||Q)&\leq -(q_{0}+\tau){\>}\ln_{\alpha}\left(\frac{q_{0}}{q_{0}+\tau}\right)
-(q_{0}-\tau){\>}\ln_{\alpha}\left(\frac{q_{0}}{q_{0}-\tau}\right)
\ , \label{lobon1}\\
H_{\alpha}(P||Q)&\leq -(q_{0}+\tau){\>}\ln_{\alpha}\left(\frac{q_{0}}{q_{0}+\tau}\right)
\ , \label{lobon2}
\end{align}
respectively for $\tau\leq{q}_{0}$ and $q_{0}<\tau\leq1-q_{0}$.
Using (\ref{alrha}) in classical setting, the bounds
(\ref{lobon1}) and (\ref{lobon2}) remain valid with
$R_{\alpha}(P||Q)$ instead of $H_{\alpha}(P||Q)$. In fact, the
function $x\mapsto\ln\bigl[1+(\alpha-1){\,}x\bigr]$ increases with
$x>0$ and $\ln(1+\xi)\leq\xi$ for $\xi\geq0$. The upper bounds
(\ref{lobon1}) and (\ref{lobon2}) are $\alpha$-parametric
extensions of the bounds obtained  for the standard relative
entropy in \cite{AE05}.

\section{Notes on the Fano and Fannes inequalities}\label{ngfn}

In this section, we will obtain upper bounds on the conditional
Tsallis $\alpha$-entropy for all $\alpha>0$. It is convenient to
change slightly the notation as follows. Let $X$ and $Y$ be
discrete random variables with probabilities $\{p_{X}(x)\}$ and
$\{p_{Y}(y)\}$, each supported on the $N$-point set $\mcx$. By
$p_{XY}(x,y)$ and $p_{X|Y}(x|y)$ we respectively denote the joint
and conditional probabilities. The joint $\alpha$-entropy and the
conditional $\alpha$-entropy are respectively defined as
\begin{align}
 & H_{\alpha}(X,Y):=\frac{1}{1-\alpha}\left(\sum\nolimits_{x,y}p_{XY}(x,y)^\alpha-1\right)
\ , \label{joind}\\
 & H_{\alpha}(X|Y):=\sum\nolimits_{y}p_{Y}(y)^{\alpha}H_{\alpha}(X|y)
\ , \label{cond0}
\end{align}
where
$H_{\alpha}(X|y)=(1-\alpha)^{-1}\Bigl(\sum\nolimits_{x}p_{X|Y}(x|y)^\alpha-1\Bigr)$.
Rewriting
$$
H_{\alpha}(X|y)=-\sum\nolimits_{x}p_{X|Y}(x|y)^{\alpha}\ln_{\alpha}p_{X|Y}(x|y)
\ ,
$$
we further obtain
\begin{equation}
H_{\alpha}(X|Y)=-\sum\nolimits_{x,y}p_{XY}(x,y)^{\alpha}\ln_{\alpha}p_{X|Y}(x|y)
\ , \label{rwha}
\end{equation}
due to $p_{Y}(y){\,}p_{X|Y}(x|y)=p_{XY}(x,y)$. We will
follow the original scheme of derivation (see the classical text
\cite{fano}, section 6.2). The probability of error is expressed
as
\begin{equation}
P_e=\sum\nolimits_{y}p_{Y}(y){\,}q(e|y)
\ , \qquad
q(e|y)=1-p_{X|Y}(y|y)=\sum\nolimits_{x\neq{y}}p_{X|Y}(x|y)
\ . \label{pep}
\end{equation}

\newtheorem{lem4}[lem1]{Lemma}
\begin{lem4}\label{metfan}
For all $\alpha\in(0,\infty)$, there holds
\begin{equation}
H_{\alpha}(X|Y)\leq\sum\nolimits_{y}p_{Y}(y)^{\alpha}h_{\alpha}\bigl(q(e|y)\bigr)+
\ln_{\alpha}(N-1)\sum\nolimits_{y}p_{Y}(y)^{\alpha}q(e|y)^{\alpha}
\ . \label{resub}
\end{equation}
\end{lem4}

{\bf Proof.} Using the expression for $q(e|y)$ and the
definition (\ref{hbin}), we write
\begin{align}
&H_{\alpha}(X|y)=-p_{X|Y}(y|y)^{\alpha}\ln_{\alpha}p_{X|Y}(y|y)-
\sum\nolimits_{x\neq{y}}p_{X|Y}(x|y)^{\alpha}\ln_{\alpha}p_{X|Y}(x|y)
\nonumber\\
&=h_{\alpha}\bigl(q(e|y)\bigr)+q(e|y)^{\alpha}\ln_{\alpha}q(e|y)-
\sum\nolimits_{x\neq{y}}p_{X|Y}(x|y)^{\alpha}\ln_{\alpha}p_{X|Y}(x|y)
\ . \label{conw}
\end{align}
Due to $q(e|y)=\sum_{x\neq{y}}p_{X|Y}(x|y)$ and the properties of
$\alpha$-logarithm, the second and third terms in the right-hand
side of (\ref{conw}) are combined as
\begin{align}
&-\sum\nolimits_{x\neq{y}}
p_{X|Y}(x|y)^{\alpha}\ln_{\alpha}p_{X|Y}(x|y)+q(e|y){\,}q(e|y)^{\alpha-1}\ln_{\alpha}q(e|y)\nonumber\\
&=-\sum\nolimits_{x\neq{y}}
p_{X|Y}(x|y)^{\alpha}\Bigl(\ln_{\alpha}p_{X|Y}(x|y)-p_{X|Y}(x|y)^{1-\alpha}q(e|y)^{\alpha-1}\ln_{\alpha}q(e|y)\Bigr)
\nonumber\\
&=-\sum\nolimits_{x\neq{y}}
p_{X|Y}(x|y)^{\alpha}\left(\ln_{\alpha}p_{X|Y}(x|y)+p_{X|Y}(x|y)^{1-\alpha}\ln_{\alpha}\frac{1}{q(e|y)}{\>}\right)
\label{onx}\\
&=-q(e|y)^{\alpha}\sum\nolimits_{x\neq{y}} \frac{p_{X|Y}(x|y)^{\alpha}}{q(e|y)^{\alpha}}
{\ }\ln_{\alpha}\frac{p_{X|Y}(x|y)}{q(e|y)}\leq
q(e|y)^{\alpha}\ln_{\alpha}(N-1)
\ . \label{stn1}
\end{align}
Here we used the identities
$\ln_{\alpha}(1/z)=-z^{\alpha-1}\ln_{\alpha}z$ (right before
(\ref{onx})) and
$\ln_{\alpha}(\xi{z})=\ln_{\alpha}\xi+\xi^{1-\alpha}\ln_{\alpha}z$
(right before (\ref{stn1})). Substituting (\ref{stn1}) in
(\ref{conw}) and further in (\ref{cond0}), we obtain
(\ref{resub}). $\blacksquare$

\newtheorem{tem3}[tem1]{Theorem}
\begin{tem3}\label{tbow}
Let random variables $X$ and $Y$ take values on the same finite
set of cardinality $N$. For given value of the error probability
$P_e$, the conditional entropy $H_{\alpha}(X|Y)$ is bounded from
above as
\begin{align}
H_{\alpha}(X|Y) &\leq\frac{P_{e}^{{\,}\alpha}-\alpha{P}_{e}}{1-\alpha}+P_{e}^{{\,}\alpha}\ln_{\alpha}\bigl[N(N-1)\bigr]
 & (0<\alpha<1)
\ , \label{fanin0}\\
H_{\alpha}(X|Y) &\leq h_{\alpha}(P_{e})+P_{e}^{{\,}\alpha}\ln_{\alpha}(N-1)
 & (1<\alpha<\infty)
\ . \label{fanin1}
\end{align}
\end{tem3}

{\bf Proof.} For $\alpha\in(0,1)$, we use the formula
$h_{\alpha}(u)=(1-\alpha)^{-1}[u^{\alpha}+(1-u)^{\alpha}-1]\leq(1-\alpha)^{-1}(u^{\alpha}-\alpha{u})$,
which follows from (\ref{hbin}) and the inequality
\begin{equation}
1-(1-u)^{\alpha}=\int\nolimits_0^u\alpha(1-t)^{\alpha-1}dt
\geq\int\nolimits_0^u\alpha{\,}dt=\alpha{u}
\ . \label{inu11}
\end{equation}
By these relations and $\xi_{y}=p_{Y}(y){\,}q(e|y)$, the first sum
in the right-hand side of (\ref{resub}) is no greater than
\begin{equation}
\frac{1}{1-\alpha}{\>}\sum\nolimits_{y} p_{Y}(y)^{\alpha}\bigl[q(e|y)^{\alpha}-\alpha{q}(e|y)\bigr]
\leq\frac{1}{1-\alpha}\left(\sum\nolimits_{y} \xi_{y}^{\alpha}-\alpha{P}_{e}\right)
\ , \label{xypal}
\end{equation}
since
$\sum_{y}p_{Y}(y)^{\alpha}q(e|y)\geq\sum_{y}p_{Y}(y){\,}q(e|y)=P_{e}$.
Using the H\"older inequality, we also obtain
\begin{equation}
\max\left\{\sum\nolimits_{y=1}^{N} \xi_y^{\alpha}:{\ }0\leq\xi_y\leq1,{\ }
\sum\nolimits_{y=1}^{N}\xi_{y}=P_{e}\right\}=N^{1-\alpha}P_{e}^{{\,}\alpha}
\ , \label{funx}
\end{equation}
which is reached for $\xi_{y}=P_{e}/N$. So the term
$(1-\alpha)^{-1}\left(N^{1-\alpha}P_{e}^{{\,}\alpha}-\alpha{P}_{e}\right)$
is an upper bound on the right-hand side of (\ref{xypal}).
Adding this with the product of $\ln_{\alpha}(N-1)$ and
(\ref{funx}) finally gives (\ref{fanin0}).

Using $p_{Y}(y)^{\alpha}\leq{p}_{Y}(y)$ for $\alpha>1$ and Jensen's
inequality for the concave function (\ref{hbin}), we have
\begin{equation}
\sum\nolimits_{y}p_{Y}(y)^{\alpha}h_{\alpha}\bigl(q(e|y)\bigr)\leq
\sum\nolimits_{y}p_{Y}(y){\,}h_{\alpha}\bigl(q(e|y)\bigr)\leq
h_{\alpha}\left(\sum\nolimits_{y}p_{Y}(y){\,}q(e|y)\right)=h_{\alpha}(P_{e})
\label{hconc}
\end{equation}
and also $\sum\nolimits_{y}
p_{Y}(y)^{\alpha}q(e|y)^{\alpha}\leq\left(\sum\nolimits_{y}
p_{Y}(y){\,}q(e|y)\right)^{\alpha}=P_{e}^{{\,}\alpha}$. By these
two points, the inequality (\ref{resub}) immediately leads to
(\ref{fanin1}). $\blacksquare$

For $\alpha>1$, the inequality (\ref{fanin1}) with $P_{e}$ instead
of $P_{e}^{\alpha}$ was derived in \cite{sf06}. So we obtain an
improvement of the known result. The formula (\ref{fanin0}) for
$0<\alpha<1$ is a new bound. By construction, the bound
(\ref{fanin0}) is not sharp. Nevertheless, this inequality is
sufficiently exact for small values of $P_{e}$. The bounds
(\ref{fanin0}) and (\ref{fanin1}) both show that $P_{e}\to0$
implies $H_{\alpha}(X|Y)\to0$. On the other hand, if
$H_{\alpha}(X|Y)$ is large then the probability of making an error
must be large as well. In this regard, the essence of our
inequalities concurs with a typical use of the standard Fano
inequality.

Uniform continuity is an important property of the von Neumann
entropy. The first result in this issue was given by Fannes
\cite{fannes}. The Tsallis entropy itself \cite{yanagi,zhang} and
its partial sums \cite{rast1023} also obey the continuity
property. Due to the classical Fano inequality, one can sharpen
Fannes' bound (see theorem 3.8 and its proof of Csisz\'{a}r in
\cite{petz08}). We shall now show that the Fano type inequalities
(\ref{fanin0}) and (\ref{fanin1}) lead to the Fannes inequality in
terms of Tsallis entropies. Using properties of the
$\alpha$-logarithm, the joint entropy (\ref{joind}) can be
rewritten as \cite{sf06}
\begin{equation}
H_{\alpha}(X,Y)=H_{\alpha}(X)+H_{\alpha}(Y|X)=H_{\alpha}(Y)+H_{\alpha}(X|Y)
\ . \label{hmwh}
\end{equation}
Due to $H_{\alpha}(Y|X)\geq0$, the difference
$H_{\alpha}(X)-H_{\alpha}(Y)\leq{H}_{\alpha}(X|Y)$ is bounded from
above by the right-hand side of (\ref{fanin0}) for
$\alpha\in(0,1)$ and by the right-hand side of (\ref{fanin1}) for
$\alpha\in(1,+\infty)$. For given distributions
$\{p_{X}(x)\}$ and $\{p_{Y}(y)\}$, the joint probability mass
function $p_{XY}(x,y)$ can be built in such a way that
\begin{equation}
P_{e}=D(X,Y)=\frac{1}{2}{\,}\sum\nolimits_{x}\bigl|p_{X}(x)-p_{Y}(x)\bigr|
\ . \label{pedf}
\end{equation}
This follows from the coupling inequality (see, e.g., the book
\cite{lindvall}). Setting $\{p_{X}(x)\}=\spc(\bro)$ and
$\{p_{Y}(y)\}=\spc(\bsg)$, we then have
$D(X,Y)\leq\rd(\bro,\bsg)$ (see, e.g., lemma 11.1 in
\cite{petz08}). When $N\geq2$, the right-hand side of
(\ref{fanin0}) increases with $P_{e}$ for all $0\leq{P}_{e}\leq1$, the
right-hand side of (\ref{fanin1}) increases with $P_{e}$ for all
$0\leq{P}_{e}\leq(N-1)/N$. Replacing $D(X,Y)$ with larger
$\rd(\bro,\bsg)$, we obtain the following result.

\newtheorem{tem4}[tem1]{Theorem}
\begin{tem4}\label{ffnn}
Let $d$ be dimensionality of the Hilbert space and
$\tau=\rd(\bro,\bsg)$; then
\begin{align}
\bigl|\rh_{\alpha}(\bro)-\rh_{\alpha}(\bsg)\bigr| & \leq
\frac{\tau^{\alpha}-\alpha\tau}{1-\alpha}+\tau^{\alpha}\ln_{\alpha}\bigl[d(d-1)\bigr]
 & (0<\alpha<1 \ , \quad 0\leq\tau\leq1)
\ , \label{hmwh20}\\
\bigl|\rh_{\alpha}(\bro)-\rh_{\alpha}(\bsg)\bigr| & \leq
h_{\alpha}(\tau)+\tau^{\alpha}\ln_{\alpha}(d-1)
 & \left(1<\alpha \ , \quad 0\leq\tau\leq\frac{d-1}{d}\right)
\ . \label{hmwh21}
\end{align}
\end{tem4}

The relation (\ref{hmwh21}), when $\alpha>1$, is just the uniform
estimate obtained by a direct method in \cite{zhang}.
Incidentally, this method allow to derive the bound for all
$0\leq\tau\leq1$. In the limit $\alpha\to1$, the inequality
(\ref{hmwh21}) reproduces the statement of theorem 3.8 in
\cite{petz08}. For $\alpha\in(0,1)$, there also exists an
inequality
\begin{equation}
\bigl|\rh_{\alpha}(\bro)-\rh_{\alpha}(\bsg)\bigr|
\leq\frac{(2\tau)^{\alpha}-2\tau}{1-\alpha}+(2\tau)^{\alpha}\ln_{\alpha}d
\ , \label{hmwh3}
\end{equation}
provided that $\|\bro-\bsg\|_1=2\tau\leq\alpha^{1/(1-\alpha)}$.
The bound (\ref{hmwh3}) was actually proved in \cite{yanagi} for
all $\alpha\in[0,2]$, but the bound (\ref{hmwh21}) is better for
$\alpha\geq1$. Comparing (\ref{hmwh20}) with (\ref{hmwh3}), we see
the following. In general, the bound (\ref{hmwh20}) is weaker but
covers all acceptable values $\tau\in[0,1]$ of the trace distance.
The scope of (\ref{hmwh3}) is restricted to the range
$0\leq2\tau\leq\alpha^{1/(1-\alpha)}$. In low dimensions, however,
the bound (\ref{hmwh20}) can be better than (\ref{hmwh3}). Say,
for $d=2$ and $\alpha=1/2$ the bound (\ref{hmwh3}) holds for
$0\leq\tau\leq1/8$. In this range, the right-hand side of
(\ref{hmwh3}) is larger than the right-hand side of
(\ref{hmwh20}). Moreover, for sufficiently small $\tau$ the
difference between the two bounds is up to $40$ \%. Thus, the
bound (\ref{hmwh20}) has some practical interest, at least in the
primary qubit case.

\acknowledgments
The author thanks anonymous referee for very detailed comments.

\end{document}